\documentclass[12pt]{article}
\usepackage{graphicx,amsmath,amssymb,color,bm}
\usepackage{amsmath,amssymb,amsthm,color}
\usepackage{graphicx}
\usepackage{cite}
\usepackage{epsfig}
\usepackage{lineno}

\renewcommand{\baselinestretch}{2}

\begin{document}
\centerline {\Large\textbf {Electric-field-induced rich magneto-absorption spectra  }}
\centerline {\Large\textbf {of ABC-stacked trilayer graphene}}
\centerline{ Yi-Ping Lin, Chiun-Yan Lin, and Min-Fa Lin$^{*}$}
\centerline{Department of Physics, National Cheng Kung University, Tainan, Taiwan}

\vskip0.6 truecm
\noindent

\begin{abstract}
The magneto-optical spectra of ABC-stacked trilayer graphene are enriched by the electric field.
A lot of prominent absorption peaks, which arise from the inter-Lnadau-level transitions, gradually change from the twin-peak structures into the double-peak ones with the increasing electric-field strength.
This comes from the destruction in mirror symmetry of xy-plane and the non-equivalence for two sublattices with the identical projections.
Specially, a single threshold peak becomes a double-peak structure, owing to the Fermi-Dirac distribution.
The absorption frequencies continuously grow or decline except for the anti-crossings of Landau levels.
Also, such anti-crossings can induce extra double-peak structures.

\end{abstract}

\vskip0.6 truecm
$\mathit{PACS}$: 81.05.U-, 78.67.Pt, 71.70.Di

\pagebreak
\renewcommand{\baselinestretch}{2}
\newpage

\vskip 0.6 truecm

%%%MAIN TEXT%%%%
Graphene, with a fascinating hexagonal lattice structure, was first fabricated by mechanical exfoliation in 2004.\cite{novoselov2004electric(graphene-exitence)}
Its unique electronic and optical properties can be easily modulated by changing the layer number \cite{Electronic-mutilayer}, the stacking configuration \cite{E-field-multilayer}, the external magnetic \cite{goerbig2011electronic} and electric field \cite{castro2007biased}, and the deformed structure \cite{wong2012strain}.
These geometric varieties and tuning effects have attracted numerous theoretical and experimental studies especially for the magneto-electronic and -optical properties in few-layer graphenes.\cite{McCann2006landau,morimoto2013theory,moon2012energy}
The quantized Landau levels (LLs) are strongly dependent on the stacking configuration.
An N-layer graphene has N groups of conduction and valence LLs.
The three highly symmetric layer configurations, AAA-, ABA-, and ABC-stacked trilayer graphene, exhibit the totally different magneto-optical properties.
The three categories of excitation channels, belonging to intra-group optical transistions, are revealed in the AA-stacked trilayer graphene.
The inter-group LL excitations are absent because of the special relationships among the LL wave functions.\cite{AAA-no_F}
The ABA trilayer one is expected to present five categories of absorption peaks since their LLs can be regarded as the superposition of those of the AB-stacked bilayer and monolayer systems.
Moreover, the ABC-stacked trilayer graphene possesses nine categories of absorption peaks, including all the intra-group and inter-group LL excitations.\cite{YP-second}
A specific magneto-optical selection rule, $\Delta n =1$ in the variation of quantum number, is identified in the above-mentioned systems.
In particular, there are three kinds of the selection rules, $\Delta n = $0, 1, and 2, in the sliding graphene with various stacking configurations.\cite{Yao-Kung2014twist}
On the experimental side, only the intra-group excitation channels from the first LL group in the AB-stacked few-layer graphenes (N=1 $\sim$ 5) have been verified until now.\cite{ABA-France}
In this work, we focus on how a uniform perpendicular electric field ($\mathbf{E}=E_{0}\hat{z}$) affects the low-frequency magneto-optical properties in the ABC-stacked trilayer graphene, including the structure, intensity and frequency of absorption peaks.

We have developed the generalized tight-binding model for various external fields \cite{HO.J.H-monolayer-PhysicaE-2008,modulated-electric-field}, in which the Hamiltonian matrix is built from the tight-binding basis functions, i.e. the subenvelope functions on the different sublattices.
It can be further utilized to investigate the magneto-optical properties of ABC-stacked trilayer graphene in $\mathbf{E}$.
Detailed investigations have been performed on how the inter-LL optical transitions are diversified by the relationship between the interlayer atomic interactions and the electric field.
This study shows that the electric field causes the splitting of Landau levels, and hence the abundant absorption spectra are induced, such as the $E_{0}$-dependent peak structure, frequency and intensity.
The twin-peak structure or a single-peak structure will be replaced by a double-peak structure at a sufficiently large electric-field strength.
The extra peaks are generated by the intragroup LL anticrossings during the variation of $E_{0}$; futhermore, nduced by the abnormal $E_{0}$-dependent absorption frequencies are obtained.
The theoretical predictions could be verified by the optical spectroscopy methods.\cite{Raman-1,infrared-1,infrared-2,Raman-magnetic}

The magnetic field, electric field and interlayer atomic interactions are simultaneously taken into account in the generalized tight-binding model.
The primitive unit cell of the ABC trilayer graphene consists of six carbon atoms ($A^{1}$, $B^{1}$, $A^{2}$, $B^{2}$, $A^{3}$; $B^{3}$), where superscripts 1, 2, and 3 correspond to the first, second, and third layer.
The atomic interactions include one intralayer type ($\beta_{0}=3.16$ eV) and five interlayer types ($\beta_{0}=3.16$ eV, $\beta_{1}=0.36$ eV, $\beta_{2}=-0.01$ eV, $\beta_{3}=0.32$ eV, $\beta_{4}=0.03$ eV; $\beta_{5}=0.013$ eV).
The vector potential caused by a uniform perpendicular magnetic field introduces a periodic phase in the atomic interactions.
Thus, the unit cell is enlarged by $2R_{B}$ times ($R_{B}=79000/B_{0}$), and the Hamiltonian is a $12R_{B} \times 12R_{B}$ Hermitian matrix.
Moreover, a uniform perpendicular electric field leads to a potential energy difference ($V_{g}=E_{0}d$; interlayer distance d=3.35 $\AA$) between two neighboring layers, i.e., $V_{g}$ changes the site energies of the diagonal Hamiltonian matrix elements.
The eigenvalues and eigenfunctions can be solved effectively by a special diagonalizing method which is used to obtain a band-like Hamiltonian matrix.

The optical absorption function according to the Fermi golden rule at zero temperature is directly obtained through
%\begin{widetext}
\begin{eqnarray*}
A(\omega)\propto\sum\limits_{c,v,n,n'}\int_{1stBZ}\frac{d\mathbf{k}}{(2\pi)^2}\mid\langle\Psi^{c}(\mathbf{k},n)|\frac{\hat{\mathbf{E}}\cdot\mathbf{P}}{m_{e}}|\Psi^{v}(\mathbf{k},n')\rangle\mid^{2}
\times Im[\frac{f(E^{c}(\mathbf{k},n))-f(E^{v}(\mathbf{k},n'))}{E^{c}(\mathbf{k},n)-E^{v}(\mathbf{k},n')-\omega-i\Gamma}],
\end{eqnarray*}
%\end{widetext}
where the superscripts c and v represent, respectively, a conduction and a valence state, $f(E^{c,v}(\mathbf{k},n))$ is the Fermi-Dirac distribution function, and $\Gamma$ ($\simeq1$ meV) is the broadening parameter.
The absorption spectrum accounts for the inter-LL transitions from the occupied state to the unoccupied one.
Only vertical transitions of $\bigtriangleup \mathbf{k_{x}}=0$ and $\bigtriangleup \mathbf{k_{y}}=0$ are allowed beacuse of the zero photon momentum.
The absorption intensity is dominated by the velocity matrix, which is calculated by the gradient approximation.\cite{MFLIN1994}
For the electric polarization $\hat{\mathbf{E}} \parallel \hat{y}$, the velocity matrix elements are obtained from the multiplication of the three matrices corresponding to the initial state, the final state and $\partial H/\partial k_{y}$.
The spectral intensity is dominated by the nearest-neighbor  Hamiltonian matrix elements, since the in-plane atomic interaction $\beta_{0}$ makes the largest contribution to the inter-LL excitations.
The optical transistions are available only when the subenvelope function of the $A^{i}$ ($B^{i}$)sublattice in the initial state has the same mode with that of $B^{i}$ ($A^{i}$) sublattice in the finial state.

In the presence of an electric field, each LL is twofold degenerate rather than the ordinary fourfold degenerate, mainly owing to the destruction in xy-plane mirror symmetry.
The six subenvelope functions are localized at 2/6 (5/6; purple) and 4/6 (1/6; blue) positions, as shown in Fig. 1 for $B_{0}=48$ T and $V_{g}=0.23$ eV.
The quantum numbers of the first LL group, $n_{1\beta}^{c,v}$ and $n_{1\alpha}^{c,v}$, are characterized by the $B_{j}^{1}$ and $A_{j}^{3}$ sublattices, respectively, where the subscripts $\beta$ and $\alpha$ stand for the localized positions 2/6 and 4/6.
Even with the same quantum number, the conduction and the valence LLs have their six subenvelope functions behave differently.
For example, the amplitudes of the $A_{j}^{1}$ and $A_{j}^{2}$ ($B_{j}^{2}$ and $B_{j}^{3}$) in $n_{1\beta}^{v} = 4$ ($n_{1\beta}^{c} = 4$) LL are much smaller than those in $n_{1\beta}^{c} = 4$ ($n_{1\beta}^{v} = 4$) one.
In general, the subenvelope functions corresponding to the identical planar projections, ($A_{j}^{1} \leftrightarrow B_{j}^{2}$) and ($A_{j}^{2} \leftrightarrow B_{j}^{3}$), are similar in their node numbers and amplitudes.
However, the potential difference will destroy the stacking symmetry between these two sublattices, so that the subenvelope functions are no longer equivalent for a sufficiently large $V_{g}$.
The sharp contrast of the amplitude in subenvelope functions is expected to induce dramatic changes in the absorption peaks.

The electric field and the atomic interactions lead to a feature-rich energy spectrum, as shown in Fig. 2(a).
Some LL energies ($E^{c,v}$) monotonously increase or decrease with the rising $V_{g}$, and the others demonstrate non-monotonic $V_{g}$-dependences.
In particular, the intragroup LL anticrossings occur frequently during the evolution of the $V_{g}$-dependent LL energies (black circles), since some LLs possess multi-mode at certain $V_{g}$ ranges.
The amplitudes of main and side modes in the subenvelope functions have a drastic change.\cite{Yao-Kung2014twist}
For example, the subenvelope functions, ($n_{1\alpha}^{c}=1$ \& $n_{1\alpha}^{c}=4$) and ($n_{1\beta}^{v}=1$ \& $n_{1\beta}^{v}=4$), are strongly hybridized around the anticrossing point ($V_{g}\sim$ 0.19 eV), the progressions are, respectively, illustrated in Fig. 2(b) and 2(c) for 0.15 eV ${\le\,V_{g}\le\,}$ 0.25 eV.
After that, their quantum numbers are exchanged.
The modes of the subenvelope wavefunctions become well-defined again in the further increase of $V_{g}$.
These variations in the amplitude, waveform, and zero-point number of the subenvelope function, which are modulated by $V_{g}$, directly reflect on the absorption spectra.

The low-frequency absorption spectra with respect to the first group of LLs exhibit many more special peak structures as being diversified by the electric field.
The threshold inter-LL excitation, coming from the optical transition: the occupied $n^{v}_{1}=3$ LL to the unoccupied $n^{c}_{1}=2$ LL, exhibits a single-peak structure for $V_{g}=0$, which is attributed to the Fermi-Dirac distribution and the selection rule of  $\Delta n=1$.\cite{YP-second}
The LLs with the different localization positions make the same contribution.
The LL degeneracy is destroyed by $V_{g}$; therefore, the threshold excitation becomes a double-peak structure.
The lower- and the higher-frequency comes from $n^{v}_{1\beta}=2 \to n^{c}_{1\beta}=3$ and $n^{v}_{1\alpha}=3 \to n^{c}_{1\alpha}=2$ (arrows in Fig. 2(a)) with their subenvelope functions localized at 2/6 and 4/6 positions, respectively.
The peak structure stays unchanged, and their frequencies increase with the increment of $V_{g}$.
However, the other absorption peaks for $V_{g}=0$ possess the twin-peak structures with the similar absorption intensity,\cite{YP-second} which is caused by the asymmetric energy spectrum between the occupied and unoccupied LLs.
Such structures also emerge in two different localiztion positions, 2/6 and 4/6 for a small $V_{g}$.
As denoted by (\underline{$3_{\beta}$}$4_{\beta}$ \& \underline{$4_{\beta}$}$3_{\beta}$) and (\underline{$4_{\alpha}$}$3_{\alpha}$ \& \underline{$3_{\alpha}$}$4_{\alpha}$) in Fig. 3,  two twin-peak structures or four absorption peaks are displayed at $V_{g}=0.01$ eV.
With the further increasing $V_{g}$, they are replaced by a double-peak structure (black circle at $V_{g}=0.11$ eV) since a certain peak of the twin-peak structure is almot vanishing.
In other words, the two peaks in a double-peak structure, respectively, arise from the inter-LL transitions with the distinct localization centers of 2/6 and 4/6.
Apparently, the dramatic changes in peak structure indicate the strong competition between the interlayer atomic interactions and the electric field.

The $V_{g}$-induced rich absorption spectra deserve a closer investigation.
For the equivalent sublattices, ($A_{j}^{1} \leftrightarrow B_{j}^{2}$) and ($A_{j}^{2} \leftrightarrow B_{j}^{3}$), with the identical planar projections possess the similar waveforms and amplitudes in the subenvelope functions.\cite{YP-second}
The equivalence will be destroyed when a electronic static gate voltage induces a potential energy difference on the distinct layers.
At a sufficiently large $V_{g}$, the amplitudes of $A^{1}$ and  $A^{2}$ in the valence LLs and those of $B^{2}$ and $B^{3}$ in the conduction LLs become very small (red rectangles in Fig. 1).
And the spectral intensity of the absorption peak \underline{$3_{\beta}$}$4_{\beta}$ (\underline{$4_{\alpha}$}$3_{\alpha}$) is proportional to the inner product calculated from the $B^{i}$ sublattices of $n^{v}_{\beta}=3$ ($n^{v}_{\alpha}=4$) and the $A^{i}$ sublattices of $n^{c}_{\beta}=4$ ($n^{c}_{\alpha}=3$).
However, for the absorption peak \underline{$4_{\beta}$}$3_{\beta}$ (\underline{$3_{\alpha}$}$4_{\alpha}$),
the intensity is determined by the inner product of the $A^{i}$ sublattices in $n^{v}_{\beta}=4$ ($n^{v}_{\alpha}=3$) and the $B^{i}$ sublattices in $n^{c}_{\beta}=3$ ($n^{c}_{\alpha}=4$).
As a result, the latter is much weaker than the former.

The electric field induces the intragroup LL anticrossings and thus extra double-peak structures.
The existence of extra absorption peaks is mainly controlled by the weight of the side mode in the progressive LL.
For example, the spectral intensities of \underline{$1_{\beta}$}$5_{\beta}$ and \underline{$5_{\alpha}$}$1_{\alpha}$ (brown circles in Fig. 3) begin their gradual rises from $V_{g}=0.15$ eV and reach their maximums when $V_{g}$ gets to 0.19 eV, since the subenvelope functions of $n^{v(c)}_{1\beta(\alpha)}$ =1 and $n^{v(c)}_{1\beta(\alpha)}$ =4 are strongly hybridized with each other (Fig. 2(b) and (c)).
After that, the intensities drop low under the increasing $V_{g}$ and disappear when the quantum numbers of subenvelope functions restore to be well-defined again.

The low-lying inter-LL absorption frequencies ($\omega_{\underline{n}n}$) strongly depend on the electric-field strength.
The absorption frequencies of the double-peak structures monotonously decrease as $V_{g}$ increases, while the opposite is true for the threshold double-peak ones, as shown in Fig. 4.
Such dependences are revealed to originate from the $V_{g}$-dependent LL energies.
The perpendicular external electric field induces the opening of energy gap; therefore, the absolute energies of the six valence and conduction LLs respectively located at the 2/6 and 4/6 positions, which are near the Fermi level and designated to the quantum numbers 0, 1 and 2, will rise with the increasing $V_{g}$.
However, the behavior of the others LLs at the 2/6 (4/6) position is just opposite to these six LLs.
As a result, the energy differences of the LLs can be clearly reflected on $\omega_{\underline{n}n}$.
Moreover, $\omega_{\underline{n}n}$ exhibits some discontinuous $V_{g}$-dependent structures at a certain electric field strength (black circles), mainly owing to the intragroup LL anticrossing.
The aforementioned characteristics of the $V_{g}$-induced magneto-absorption spectra, like the peak structure, peak intensity, and $V_{g}$-dependent absorption frequency, could be verified by optical spectroscopy methods, such as transmission \cite{transmission-1,multi-transmission1}, reflection \cite{reflection-1,reflection-2} and Raman scattering spectroscopies.\cite{Raman-1,Raman-trilayer,Raman-bilayer}

The generalized tight-binding model is developed to investigate optical properties of ABC-stacked trilayer graphene in magnetic and electric fields.
The prominent LL magneto-absorption peaks are diversified by the electric field.
With the increased $V_{g}$, the twin-peak structures gradually change into the double-peak ones.
The former and the latter, which correspond to the inter-LL transitions with the same and the different localization centers, are, respectively, caused by the asymmetric LL energy spectrum and the destructions in both the mirror symmetry and equivalence of two sublattices.
The special peak structures are mainly determined by the competition between the interlayer atomic interactions and the electric-field strength.
Especially, a threshold single peak is replaced by a double-peak structure because of the Fermi-Dirac distribution.
Moreover, the anti-crossings of LLs can create extra double-peak structures and thus disrupt the dependence of peak frequency on $V_{g}$.
The experimental examinations of the $V_{g}$-induced magneto-optical spectra are useful in identifying the stacking configurations among few-layered graphene systems.

\bigskip

\bigskip

\centerline {\textbf {ACKNOWLEDGMENTS}}%

\bigskip

\bigskip

%\centerline {\Large \textbf {References}}

$^{*}$e-mail address: mflin@mail.ncku.edu.tw

\newpage

\begin{figure}[tbp]
%h=here, t=top, b=bottom, p=separate figure page
\par
\begin{center}
\leavevmode
\includegraphics[width=0.7\linewidth]{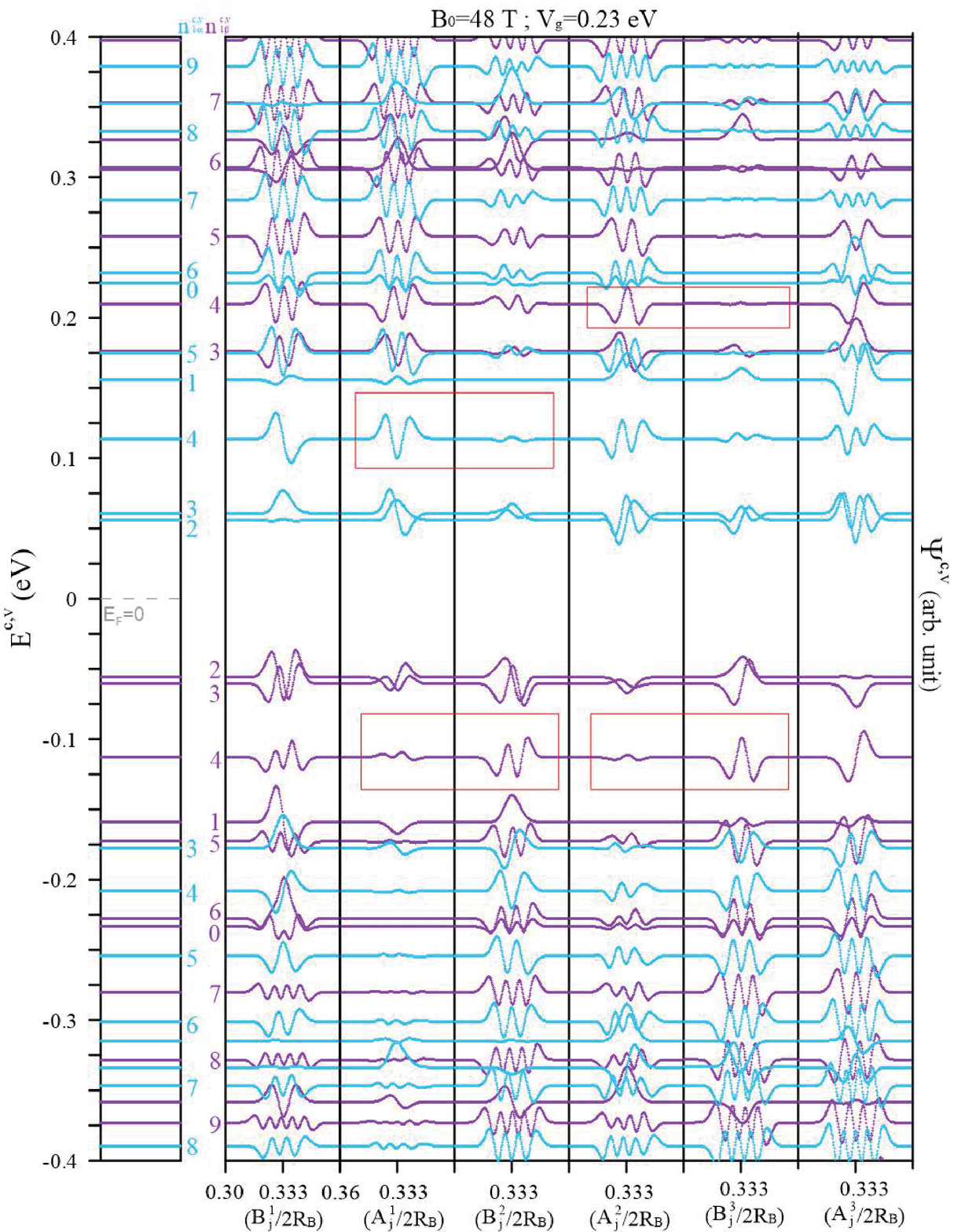}
\end{center}
\par
\textbf{Figure 1. The first group of Landau levels and their respective six subenvelope wavefunctions for $B_{0}=48$ T and $V_{g}=0.23$ eV.}
\end{figure}

\begin{figure}[tbp]
%h=here, t=top, b=bottom, p=separate figure page
\par
\begin{center}
\leavevmode
\includegraphics[width=0.7\linewidth]{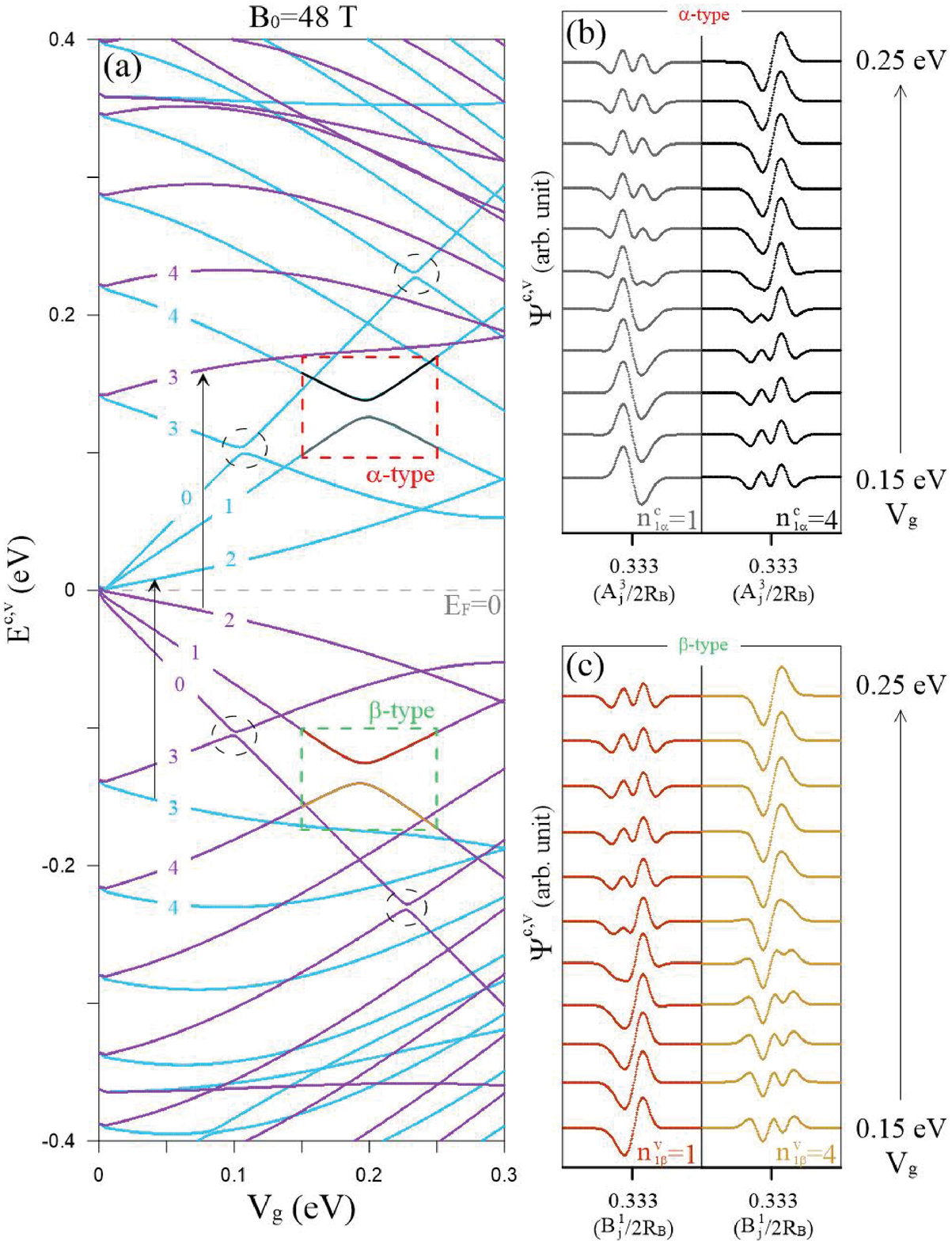}
\end{center}
\par
\textbf{Figure 2. (a) The two Landau-level subgroup spectra at $B_{0}=48$ T as the function of the electric field ($V_{g}$) illustrated in two different color codes: purple and blue. The $\alpha$-type and $\beta$-type intragroup Landau-level anticrossings are indicated by red and green rectangles, respectively. (b)The variations of the $A^{3}_{j}$ subenvelope wavefunctions for $n_{\alpha}^{c}=1$ and $n_{\alpha}^{c}=4$ with respect to the $\alpha$-type intragroup Landau-level anticrossing. (c)The variations of the $B^{1}_{j}$ subenvelope wavefunctions for $n_{\beta}^{v}=1$ and $n_{\beta}^{v}=4$ with respect to the $\beta$-type intragroup Landau-level anticrossing.}
\end{figure}

\begin{figure}[tbp]
%h=here, t=top, b=bottom, p=separate figure page
\par
\begin{center}
\leavevmode
\includegraphics[width=0.7\linewidth]{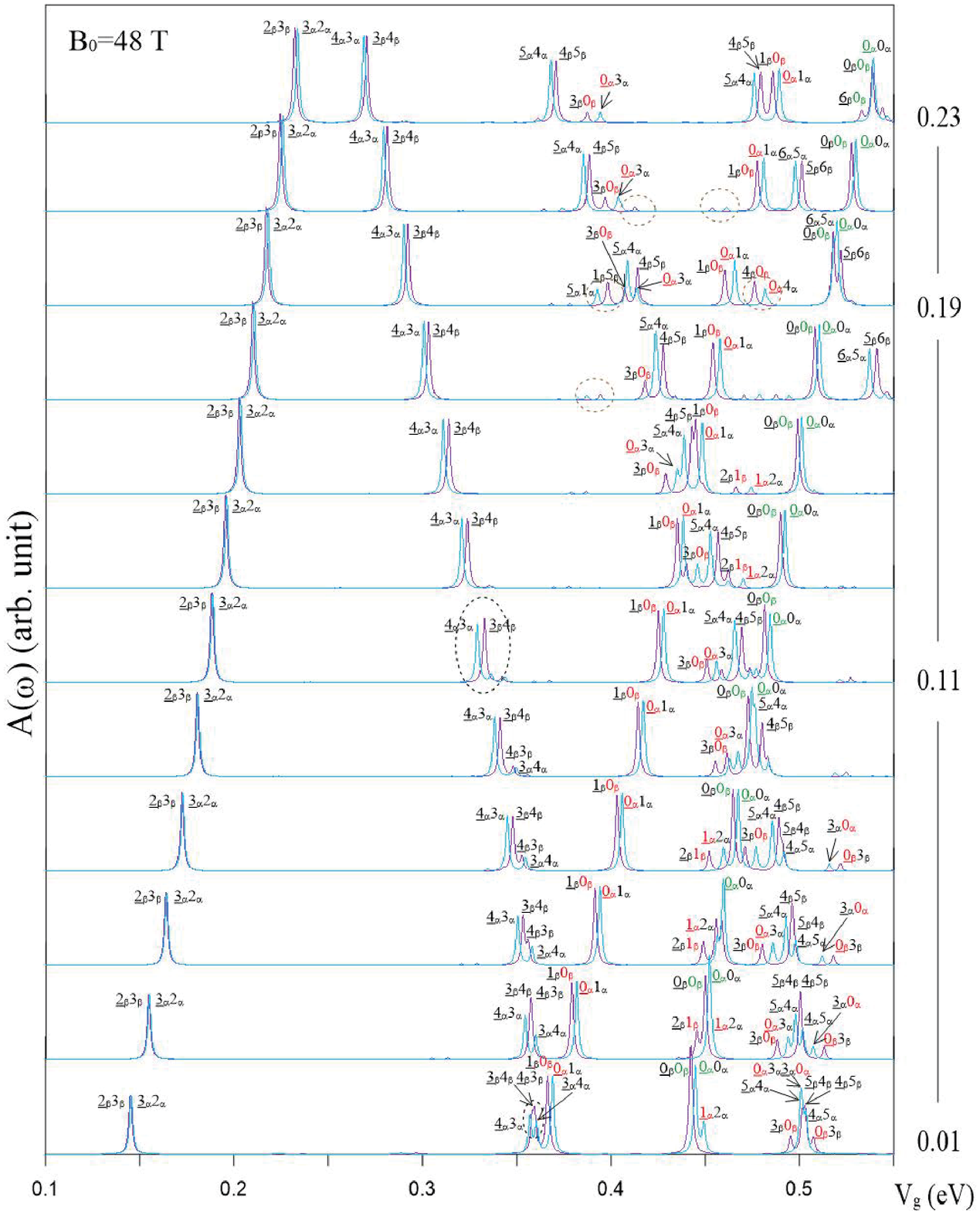}
\end{center}
\par
\textbf{Figure 3. The progression of the absorption peaks in the low-frequency region from $V_{g}=0.01$ to 0.23 eV.}
\end{figure}

\begin{figure}[tbp]
%h=here, t=top, b=bottom, p=separate figure page
\par
\begin{center}
\leavevmode
\includegraphics[width=0.7\linewidth]{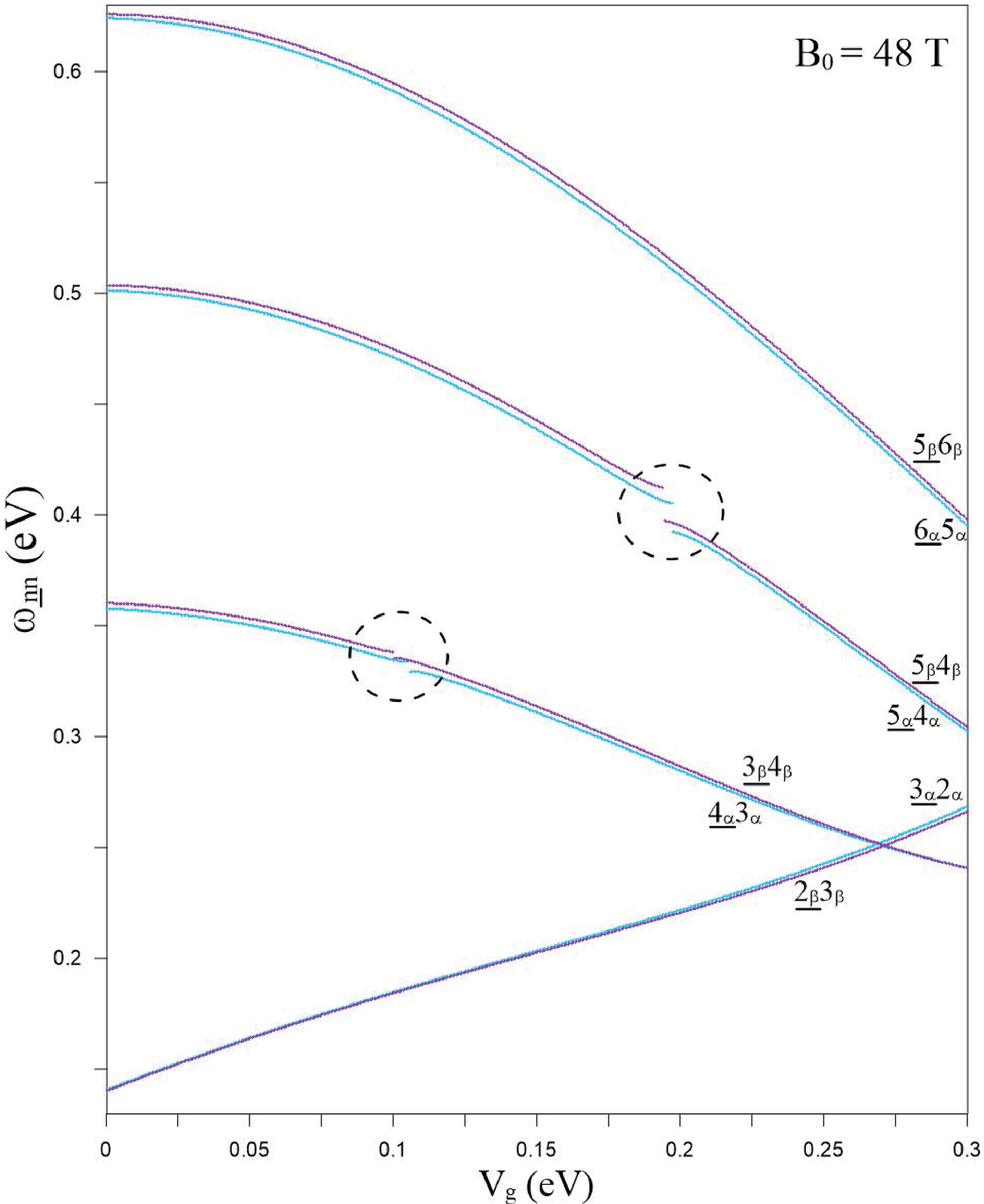}
\end{center}
\par
\textbf{Figure 4. The $V_{g}$-dependent frequencies of the low-lying inter-Landau level absorption peaks.}
\end{figure}


\newpage
\begin{thebibliography}{99}
\bibitem{novoselov2004electric(graphene-exitence)} K. S. Novoselov, A. K. Geim, S. V. Morozov, D. Jiang, Y. Zhang, S. V. Dubonos, I. V. Grigorieva and A. A. Firsov. Science 2004, 306, 666.

\bibitem{Electronic-mutilayer} J. Nilsson, A. C. Neto, F. Guinea and N. Peres. Physical Review B 2008, 78, 045405.

\bibitem{E-field-multilayer} M. Koshino. Physical Review B 2010, 81, 125304.

\bibitem{goerbig2011electronic} M. O. Goerbig. Reviews of Modern Physics 2011, 83, 1193.

\bibitem{castro2007biased} E. V. Castro, K. Novoselov, S. Morozov, N. Peres, J. L. Dos Santos, J. Nilsson,
F. Guinea, A. Geim and A. C. Neto. Physical Review Letters 2007, 99, 216802.

\bibitem{wong2012strain} J. H. Wong, B. R. Wu and M. F. Lin. The Journal of Physical Chemistry C 2012,
116, 8271.

\bibitem{McCann2006landau} E. McCann and  V. I Fal'ko,. Physical Review Letters, 2006, 96, 086805.

\bibitem{morimoto2013theory} T. Morimoto, M. Koshino and H. Aoki. Journal of Physics: Conference Series,
2013, p. 012028.

\bibitem{moon2012energy} P. Moon and M. Koshino. Physical Review B, 2012, 85, 195458.

\bibitem{AAA-no_F} C. W. Chiu, S. C. Chen, Y. C. Huang, F. L. Shyu and M. F. Lin. Applied Physics
Letters, 2013, 103, 041907.

\bibitem{YP-second} Y. P. Lin, C. Y. Lin, Y. H. Ho, T. N. Do and M. F. Lin. Phys. Chem. Chem. Phys.,
2015, 17, 15921.

\bibitem{Yao-Kung2014twist} Y. K. Huang, S. C. Chen, Y. H. Ho, C. Y. Lin and M. F. Lin.
Sci. Rep., 2014, 4,
7509.

\bibitem{ABA-France} S. Berciaud, M. Potemski and C. Faugeras. Nano letters, 2014, 14, 4548.

\bibitem{HO.J.H-monolayer-PhysicaE-2008} J. H. Ho, Y. H. Lai, Y. H. Chiu and M. F. Lin. Physica E: Low-dimensional Systems and Nanostructures, 2008, 40, 1722.

\bibitem{modulated-electric-field} S. C. Chen, T. S. Wang, C. H. Lee and M. F. Lin. Physics Letters A, 2008, 372,
5999.

\bibitem{Raman-1} A. C. Ferrari and D. M. Basko. Nature nanotechnology, 2013, 8, 235.

\bibitem{infrared-1} K. F. Mak, J. Shan and T. F. Heinz. Physical review letters, 2010, 104, 176404.

\bibitem{infrared-2} K. F. Mak, L. Ju, F. Wang and T. F. Heinz. Solid State Communications, 2012,
152, 1341.

\bibitem{Raman-magnetic} C. Cong, J. Jung, B. Cao, C. Qiu, X. Shen, A. Ferreira, S. Adam and T. Yu. Physical
Review B, 2015, 91, 235403.

\bibitem{MFLIN1994} M. F. Lin and K. W. K. Shung. Phys. Rev. B, 1994, 50, 17744.

\bibitem{transmission-1} C. Li, M. T. Cole, W. Lei, K. Qu, K. Ying, Y. Zhang, A. R. Robertson, J. H. Warner,
S. Ding, X. Zhang, et. al.  Advanced Functional Materials, 2014, 24, 1218¡V1227.

\bibitem{multi-transmission1} P. Plochocka, C. Faugeras, M. Orlita, M. Sadowski, G. Martinez, M. Potemski,
M. Goerbig, J.-N. Fuchs, C. Berger and W. De Heer. Physical review letters, 2008,
100, 087401.

\bibitem{reflection-1} Z. Yan, Y. Liu, L. Ju, Z. Peng, J. Lin, G. Wang, H. Zhou, C. Xiang, E. Samuel,
C. Kittrell, et. al. Angewandte Chemie, 2014, 126, 1591.

\bibitem{reflection-2} B. Daas, K. Daniels, T. Sudarshan and M. Chandrashekhar. Journal of Applied
Physics, 2011, 110, 113114. 
    
\bibitem{Raman-trilayer}  C. Cong, T. Yu, K. Sato, J. Shang, R. Saito, G. F. Dresselhaus and M. S. Dresselhaus. ACS nano, 2011, 5, 8760¡V8768.

\bibitem{Raman-bilayer}  K. Kim, S. Coh, L. Z. Tan, W. Regan, J. M. Yuk, E. Chatterjee, M. Crommie, M. L.
Cohen, S. G. Louie and A. Zettl. Physical review letters, 2012, 108, 246103.


\end{thebibliography}
\end{document}